\documentclass[aip,pop,reprint]{revtex4-1} 

\usepackage{graphicx}
\usepackage{amssymb,amsmath}
\usepackage{bm}
\usepackage{siunitx}
\allowdisplaybreaks

\begin{document}
\title{Effect of electron diamagnetic drifts on cylindrical double-tearing modes}
\author{Stephen Abbott}
\author{Kai Germaschewski}
\affiliation{University of New Hampshire, Durham, NH 03824}

\begin{abstract}
Double-tearing modes~(DTMs) have been proposed as a driver of `off-axis
sawtooth' crashes in reverse magnetic shear tokamak configurations. 
Recently differential rotation provided by equilibrium sheared flows has been
shown capable of decoupling the two DTM resonant layers, slowing the growth the
instability. In this work we instead supply this differential rotation using an
electron diamagnetic drift, which emerges in the presence of an equilibrium
pressure gradient and finite Larmor radius physics. Diamagnetic drifts have the
additional benefit of stabilizing reconnection local to the two tearing layers.
Conducting linear and nonlinear simulations with the extended MHD code \texttt
{MRC-3d}, we consider an $m=2$, $n=1$ cylindrical double-tearing mode. We show
that asymmetries between the resonant layers and the emergence of an ideal
MHD instability cause the DTM evolution to be highly dependent on the
location of the pressure gradient. By locating a strong drift near the outer,
dominant resonant surface are we able to saturate the mode and preserve the
annular current ring, suggesting that the appearance of DTM activity in advanced
tokamaks depends strongly on the details of the plasma pressure profile.
\end{abstract}
\maketitle

\section{Introduction}
\label{sec:introduction}
The promising properties of the reversed magnetic shear~(RMS) tokamak
configuration have led to a recent increased interest in the double tearing
mode. RMS configurations result in safety factor profiles that are
non-monotonic. If two rational surfaces of the same $q$ exist near each other
within the plasma, they may couple together via ideal MHD scale processes to form
a single double-tearing mode~(DTM). Linearly the interaction of the two surfaces
creates a self-driven reconnecting instability that depends weakly on
resistivity.\cite{Pritchett1980} Nonlinearly the DTM can
potentially disrupt the annular current ring of RMS devices,\cite{Chang1996}
generate strong sheared flows,\cite{Wang2008} and release large bursts of
kinetic energy.\cite{Ishii2000} As such, they are a proposed driver of off-axis sawtooth
behavior.

One means of stabilizing double-tearing mode activity is
the application of differential rotation to the two DTM layers. In slab
Cartesian simulations equilibrium sheared flow has been shown to interfere with
the coupling between the resonant layers to result in two decoupled, drifting
single tearing modes.\cite{Mao2013} Further increase in the flow amplitude
generates Alfv\'{e}n resonance layers that couple to the tearing surfaces,
increasing or decreasing the mode growth depending on their proximity.
Nonlinearly these Alfv\'{e}n resonances may shield the plasma core and
suppress DTM mode growth.\cite{Wang2011a,Voslion2011} The appearance of such
layers requires, however, flow amplitudes near the in-plane Alfv\'{e}n speed and
shears near the threshold for Kelvin-Helmholtz instability,\cite{Mao2013}
potentially triggering greater instability. Thus we are motivated to explore
alternate means of providing differential plasma rotation.

Diamagnetic drifts emerge as a result of
finite Larmor radius physics in the presence of a pressure gradient such as
the internal transport barriers~(ITBs) observed in RMS plasmas.\cite{Yuh2009}
They have long been studied as means of stabilizing reconnecting modes, and have
several advantages over equilibrium flow. In particular, diamagnetic drifts
local to the reconnecting
layer interfere with the conversion of magnetic energy to 
kinetic.\cite{Ara1978a} This local effect has been shown to saturate the
$m=1$ kink-tearing mode in conventional tokamaks,\cite{Rogers1995} leaving
finite sized islands during incomplete sawtooth crashes.\cite{Beidler2011} 

The influence of both pressure gradients and
diamagnetic drifts on double-tearing modes has been considered previously by
other authors. In resistive, reduced MHD simulations 
Zhao et al.~\cite{Zhao2011} examined the impact of equilibrium pressure
gradients on a cylindrical DTM with a small inter-resonant spacing. Their
results show that the pressure gradient modifies the dependence of the DTM on
resistivity, causing variations in the spectrum of modes at a given surface.
In this work we will expand their study to more widely spaced modes and a wider
variety of pressure gradients, as well as introduce finite Larmor radius
effects. Maget et al.~\cite{Maget2014} considered
neoclassical effects on a DTM in toroidal simulations, targeting Tore Supra
experiments. They found some mode numbers were stabilized by the addition of
diamagnetic drifts, whereas others were enhanced due to toroidal effects. Their
simulations target, however, specific discharges and do not consider variations
in the pressure and drift
profiles, thus do not illuminate the role of differential rotation in DTM
evolution.

In this work, we use Hall MHD simulations to examine the impact of diamagnetic
drifts on a cylindrical $m=2$, $n=1$ double-tearing mode, considering both the
ability of an electron diamagnetic drift to decouple the two tearing layers and
to stabilize them once separated. To this end, we structure this paper as
follows. In Section~\ref{sec:model} we introduce the Hall MHD model and
describe our simulation code \texttt{MRC-3d}. Section~\ref{sec:equilibrium}
defines the equilibrium safety factor and density profiles used for this study.
In Section~\ref{sec:linear} we report the results of linear resistive and Hall
MHD simulations. We find that the addition of a pressure gradient to our
equilibrium destabilizes an ideal MHD instability that competes with the
stabilizing effects of the diamagnetic drift. As a consequence, we are able to
decrease the linear DTM growth rate only by locating a strong diamagnetic drift
at the dominant, outer rational surface. We use this result to choose
characteristic profiles for nonlinear Hall MHD simulations in Section~\ref
{sec:nonlinear}, and show that the DTM may be saturated before disruption of the
annular current ring. Finally, we summarize our results in Section~\ref
{sec:conclusion} and discuss the consequence of this work for advanced tokamaks.

\section{Hall magnetohydrodynamic model}
\label{sec:model}
Our simulation code \texttt{MRC-3d} uses a standard Hall MHD model.
\begin{align}
  \label{eq1:mass}
  \frac{\partial \rho}{\partial t} &= -\mathbf{\nabla} \cdot  (\rho \mathbf{U} - D \nabla \rho)\\
  \label{eq1:momentum}
  \frac{\partial \mathbf{P}}{\partial t} &= -\mathbf{\nabla} \cdot
  [\rho\mathbf{UU} - \mathbf{BB} + \mathbf{I}(p + B^{2}/2) - \rho \nu \mathbf{\nabla U}]\\
  \label{eq1:temperature}
  \frac{\partial T_{e}}{\partial t} &= -\mathbf{U} \cdot \mathbf{\nabla}T_{e} - (\gamma - 1)T_{e} \mathbf{\nabla} \cdot \mathbf{U} \\
  \label{eq1:pressure_species}
  p_{s} &= \rho T_{s}\\
  \label{eq1:ohmslaw}
  \mathbf{E} &= -\mathbf{U} \times \mathbf{B} + \frac{d_{i}}{\rho}(\mathbf{J} \times \mathbf{B} - \nabla p_{e}) + \eta \mathbf{J}\\
  \label{eq1:faraday}
  \frac{\partial \mathbf{B}}{\partial t} &= -\mathbf{\nabla} \times \mathbf{E}\\
  \label{eq1:current}
  \mathbf{J} &= \mathbf{\nabla} \times \mathbf{B}
\end{align}
where $p = (1 + \tau) \rho T$is the pressure and $\tau = T_{i}/T_{e}$ is the
ratio of the ion to electron temperatures. In this work we focus on the cold-ion
regime ($\tau=0$), thus excluding ion diamagnetic drifts.

The simulation code \texttt{MRC-3d}\cite{mrcdocs} implements the above model in
a fully conservative, finite-volume scheme similar to Chac\`{o}n\cite{Chacon2004},
with additional $d_{i}$ scale Hall and electron pressure gradient terms.
Lengths are normalized to the cylinder radius, magnetic fields to the
asymptotic in-plane magnitude, and velocities to the in-plane Alfv\'{e}n speed.
All other normalizations follow from these. Data management and implicit
time integration are accomplished via
the PETSc~\cite{petsc-web-page,petsc-user-ref,petsc-efficient} 
interface in the \texttt{LIBMRC} computational library.\cite{libmrc}

We conduct simulations in 2D helically symmetric
cylindrical geometry. Derivatives in radial $0\leq r \leq 1$ and poloidal
$0 \leq \theta \leq 2\pi$ coordinates are discretized directly. The cylinder is
assumed to be periodic in $z$ with a length of $2\pi R$ where $R=10$ is
the major radius of an approximately equivalent torus with inverse aspect
ratio $\epsilon=0.1$. Derivatives in the axial
coordinate $\phi=z/R$ are taken to be $\mathrm{d}\phi = \iota^{-1} \mathrm{d}
\theta$ where $\iota=n/m$ is the twist of the helix. Thus for analysis we
define the helical coordinates
\begin{align}
\hat{u} &= \frac{1}{\sqrt{1 + \frac{n^{2}}{m^{2}}\frac{r^{2}}{R^{2}}}} \left [ \hat{\theta} - \frac{n}{m} \frac{r}{R} \hat{z} \right ] \\
\hat{h} &= \frac{1}{\sqrt{1 + \frac{n^{2}}{m^{2}}\frac{r^{2}}{R^{2}}}} \left [ \hat{z} + \frac{n}{m} \frac{r}{R}\hat{\theta} \right ]
\end{align}
where $\hat{r}$ and $\hat{u}$ represent the two dimensional perpendicular plane
and $\hat{h}$ is directed along the helix. The helical flux function $\psi^{*}$
is then defined as 
\begin{align}
	\bm{B} &=  \nabla \psi^{*} \times  \hat{h}+ B_{h} \hat{h}
\end{align}

To reduce the computational costs of this work we use resistivities on
the order of $\eta\sim10^{-5}$. This unrealistically large level of diffusion
causes the equilibrium to decay on a time scale comparable to the growth time of
resistive instabilities. \texttt{MRC-3d} features a mechanism to prevent
equilibrium decay that is equivalent to the addition of a source electric field
in Ohm's law~(Eqn.~\ref{eq1:ohmslaw}). In nonlinear simulations we enable this
mechanism and prevent the resistive decay of the equilibrium. We have confirmed
that the major results of this work persist when the source electric field is
disabled, and will note explicitly when it impacts the mode behavior.

\section{Equilibrium}
\label{sec:equilibrium}
To generate an equilibrium with two nearby $q=2$ rational surfaces we use the
safety factor profile given by Bierwage\cite{Bierwage2005}: 
\begin{equation}
\begin{aligned}
	\label{eq2:dtm_q_profile}
	q(r) &= q_{0} F_{1}(r) \left \{ 1 + (r/r_{0})^{2 w(r)}
    \right \}^{1/w(r)} \\
    r_{0} &= r_{A} | [m/(nq_{0})]^{w(r_{A})} - 1|^{-1/[2w(r_{A})]} \\
    w(r) &= w_{0} + w_{1}r^{2} \\
    F_{1}(r) &= 1 + f_{1} \exp \left \{ - [(r - r_{11}) /
    r_{12}]^{2} \right \}
\end{aligned}
\end{equation}
with the following constants:
\begin{equation}
\begin{aligned}
 	\label{eq2:q_profile_fixed_vals}
    r_{A} &= 0.655, & w_{0} &= 3.8824, & w_{1} &= 0\\
    f_{1} &= -0.238, &r_{11} &= 0.4286, &r_{12} &= 0.304
\end{aligned}
\end{equation}
We set $q_{0}=2.5$, resulting in two $q=2$ rational surfaces
located at $0 < r_{s1} < r_{s2} < 1$, spaced a distance $D = r_
{s2}-r_{s1} \approx0.26$ apart.

\begin{figure}
	\centering
	\includegraphics[width=\columnwidth]{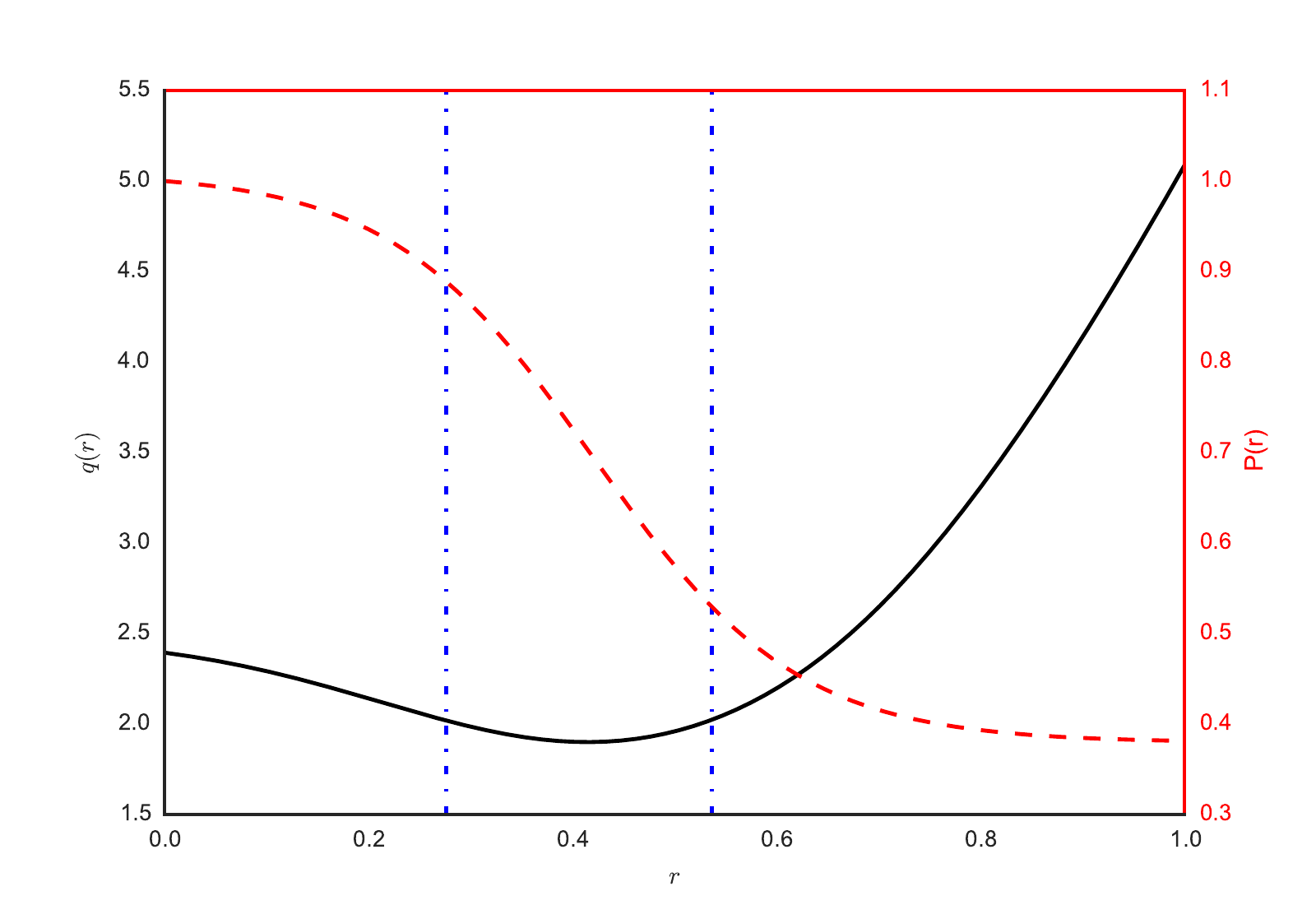}
	\caption{\label{fig:q_P_ex}Equilibrium safety factor profile (solid, left axis)
	with two nearby $q=2$ rational surfaces, indicated by dot-dashed vertical
	lines, and an example pressure profile (dashed, right axis).}
\end{figure}

Diamagnetic drifts require the introduction of a pressure gradient. We use a
monotonic density profile to represent an internal transport barrier as given
by Zhao\cite{Zhao2011}:
\begin{align}
\label{eq2:cyl_pressure_profile}
\rho(r) = N_{0} \left \{ 1 - (1 - N_{b}) \frac{\tanh
(r_{0}/\delta_{N}) + \tanh [ (r -
r_{0})/\delta_{N}]}{\tanh(r_{0}/\delta_{N}) + \tanh[(1 - r_{0}) \delta_{N}]}
\right \}
\end{align}
We fix the core density with $N_{0}=1$ and vary $r_{0}$, $\delta_{N}$, and
$N_{b}$ to change the center, width, and edge magnitude of the density profile.
For simplicity we take the equilibrium electron temperature to be a
constant $T_ {0}=1.0$, and assume cold-ions $\tau=T_{i}/T_{e}=0$ .
We initialize the equilibrium magnetic field $\bm{B}_{0}$ by specifying a
density profile and iteratively refining $B_{\theta 0}$ and $B_{z0}$ toward the
above safety factor profile,
subject to the constraints of force-balance and that $B_ {z0}\sim 10$.
This equilibrium has a plasma parameter of $\beta \approx 0.01$ so that
$\beta\sim\epsilon^{2}$, consistent with the standard tokamak ordering
assumption. The equilibrium safety factor and example density gradient are shown
in Figure~\ref{fig:q_P_ex}.

This pressure gradient produces~(in the cold ion regime with a constant
electron temperature $T_{0}=1.0$) an electron diamagnetic drift given
by:
\begin{align}
  \label{eq:drift}
  \omega_{*}(r) &= \bm{k} \cdot \bm{v}_{*e} = - \frac{m d_{i} B_{h}}{r\rho B^{2}} \frac{\partial \rho}{\partial r}
\end{align}
where we have assumed the perturbation wave vector is $\bm{k} = m/r \hat
{\theta} - n/R \hat{z}$, commensurate with our helical symmetry.

\begin{figure*}[t]
	\centering
 	\includegraphics[width=\textwidth]{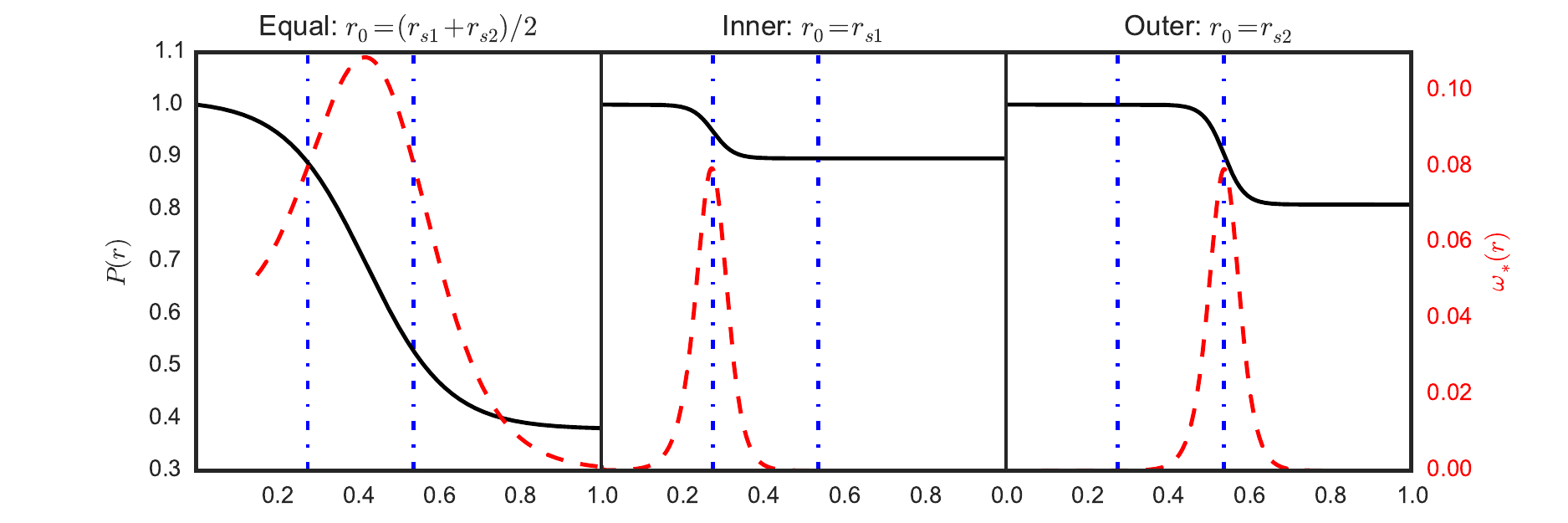}
 	\caption{\label{fig:drift_profiles}Examples of the three pressure profile
 	types (solid lines) and the electron diamagnetic drifts they produce (dashed
 	lines) from left to right: equal drift; inner drift; and outer drift. Vertical
 	dash-dot lines indicate the locations of the $q=2$ rational surfaces. All
 	three profiles results in a drift of $\omega_{*}=0.8$ at one or both of the
 	rational surfaces, respectively.}
\end{figure*}

We choose three classes of density profiles, shown in Figure~\ref
{fig:drift_profiles}, which center the maximum gradient at different locations
$r_{s1} \leq r_{0} \leq r_{s2}$.
\emph{Equal drift} profiles have the peak gradient centered between the two
$q=2$ resonant surfaces at $r_ {0}= (r_ {s1} + r_ {s2})/2$ and a broad width of
$\delta_{N}=0.2$, producing equal $\omega_ {*}$ at both singular layers.
\emph{Inner drift} profiles have a narrow pressure profile of $\delta_{N}=0.05$
centered at $r_ {0}=r_ {s1}$, providing a strong drift at the inner rational
surface and negligible $\omega_{*}$ at the outer. Finally, \emph{outer drift}
profiles are localized near $r_ {0}=r_ {s2}$ with $\delta_{N}=0.05$ so that the
inner rational surface experiences negligible $\omega_{*}$. Equal drift profiles
demonstrate the stabilizing effects of local diamagnetic drift
on both surfaces. The inner and outer drift  profiles produce a differential
diamagnetic drift $\Delta \omega_{*}=|\omega_{*}(r_{s1}) - \omega_{*}(r_{s2})|$,
resulting in an additional differential rotation effect and asymmetric local
stabilization of the two surfaces. The impact of diamagnetic drifts in more
realistic ITB-like profiles will likely be some intermediate form of these three
prototypical equilibrium types.

\section{Linear drifts}
\label{sec:linear}
The basic consequences of introducing diamagnetic drifts to DTM evolution
are clearly observable during the linear phase. \texttt{MRC-3d} includes a one
dimensional, linearized form of the model given
in Section~\ref{sec:model}, which we will use for this portion of the study.
We treat derivatives in $r$ using finite volume discretization and apply the
Fourier ansatz $F(r,\theta,z)=f(r)\exp{(m\theta/r - nz/R)}$ to
derivatives in $\theta$ and $z$, where the poloidal and
toroidal mode numbers $m=2$ and $n=1$ are chosen to capture the lowest (and
fastest growing) harmonic. From initial value simulations we fit
the growth rate $\gamma_{R}$ of the mode from the time evolution of
the magnetic and momentum field amplitudes. To extract the mode drift
frequency $\gamma_{I}$ we apply a discrete Fourier
transformation (DFT) to the time series output of the helical flux function
$\psi^{*}$.

The reduction to a one dimensional linear model allows us to easily
conduct scaling studies of DTM behavior in the three equilibria types given
above. For each profile type, we run simulations over a range of
diamagnetic drift values $0\leq \omega_ {*} \leq 0.16$. For each simulations we
specify the center of the gradient $r_ {0}$ and the desired drifts at the inner
($\omega_{*}(r_{s1})$) and outer ($\omega_{*}(r_{s2})$) $q=2$ rational surfaces,
then iteratively refine the density height $N_{b}$ and the magnetic fields to
produce the desired profile. We then seed a small $m=2$, $n=1$ perturbation onto
this equilibrium and run the initial value simulation using a resistivity 
$\eta=\num {1e-5}$.  To enhance numerical stability we add in small
amounts of viscosity, particle diffusivity, and temperature diffusivity
$\nu=D=DT=10^{-1}\times\eta$. These extra dissipation coefficients smooth noise
in the linear simulations and allow easier analysis; we have confirmed that they
have a negligible impact on the measured linear growth rates.

\subsection{Ideal MHD instability}
\begin{figure}
	\includegraphics[width=\columnwidth]{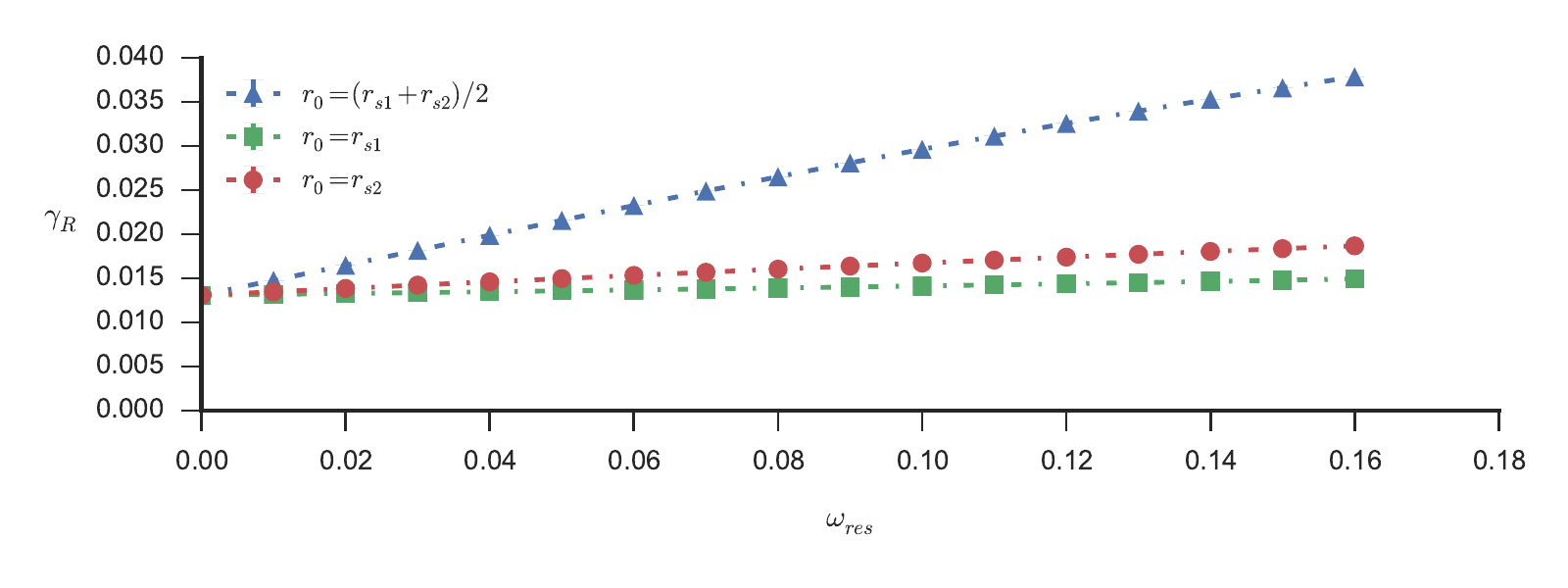}
	\caption{\label{fig:linear_res}Variation in linear $m=2$, $n=1$ DTM growth rates
  $\gamma_ {R}$ with pressure gradient in resistive MHD. For better
  comparison with Figure~\ref{fig:linear_hall} we represent the pressure
  gradients using $\omega_{res}$, which is the electron diamagnetic drift each
  profile \emph{would} produce at the specified resonant surfaces were $d_
  {i}=0.1$ rather than $0$. The three
  classes of profile described in Section~\ref{sec:equilibrium} are
  represented as: triangles--$r_ {0}= (r_{s1}+r_{s2})/2$~(equal drift);
  squares--$r_{0}=r_{s1}$~(inner drift); circles--$r_{0}=r_{s2}$~(outer drift).}
\end{figure}

Before examining the impact of the diamagnetic drift we first set $d_{i}=0$ and
consider the addition of a pressure gradient in resistive MHD. In Figure~\ref{fig:linear_res}
we have plotted the dependence of the linear growth rate $\gamma_{R}$ 
on the maximum of the pressure gradient $\partial_{r}P|_{r=r_{0}}$. Although the
diamagnetic drift is not present in these resistive simulations, we will
continue referring to the three types of profiles as 
equal~($r_{0}=(r_{s1}+r_{s2})/2$), inner~($r_{0}=r_{s1}$), and outer~($r_{0}=r_{s2}$)
`drift' configurations.

\begin{figure}
	\includegraphics[width=\columnwidth]{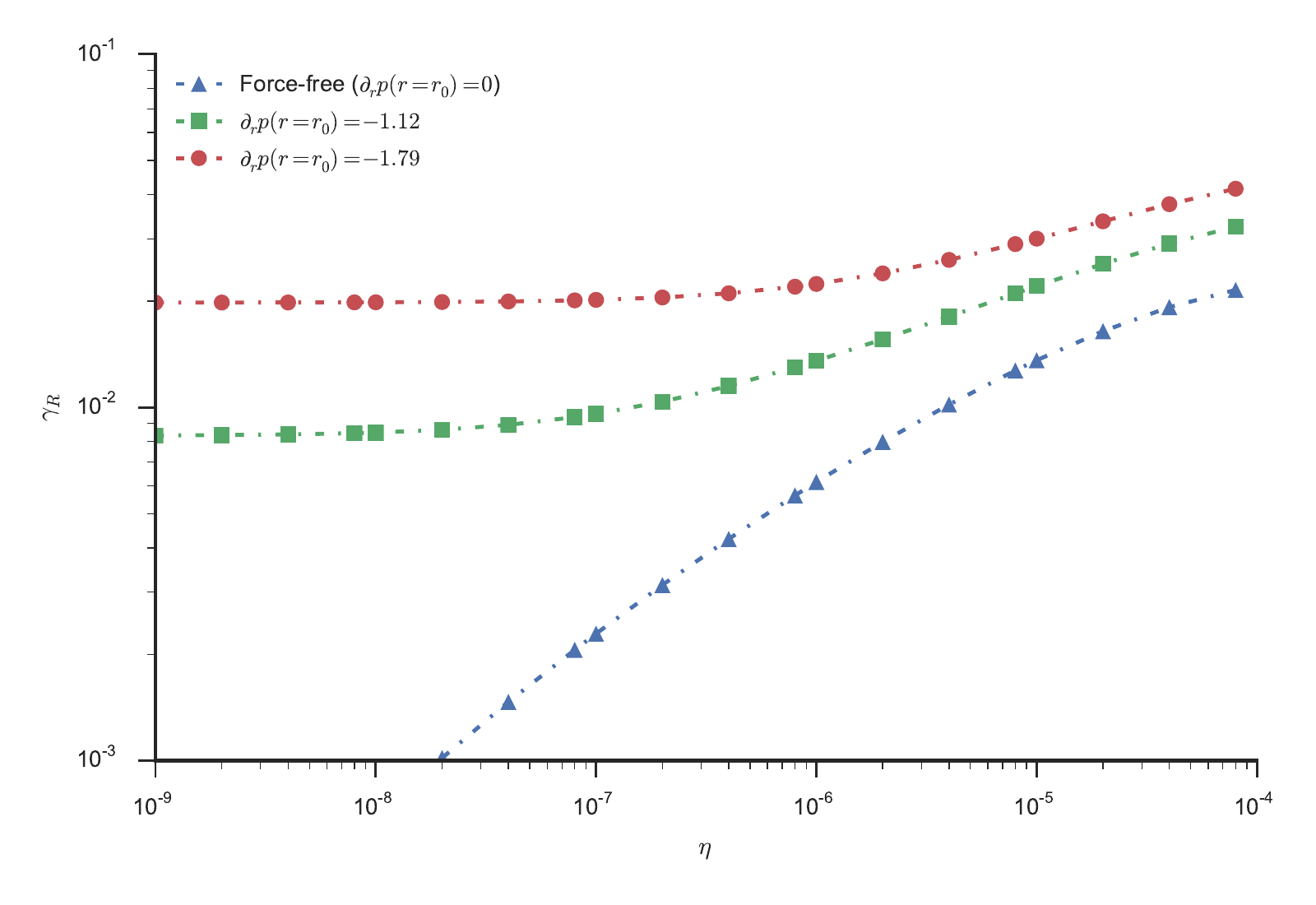}
	\caption{\label{fig:ideal_eta}Dependence of the $m=2$, $n=1$ DTM on
	resistivity $\eta$ in the presence of a pressure gradient centered at $r_ {0}=
	(r_{s1}+r_ {s2})/2$~(equal drift profile) for different values of the peak
	pressure gradient. Triangles $\partial_{r}P|_{r=r_{0}}=0$; squares $\partial_
	{r}P|_ {r=r_{0}}=-1.12$; circles $\partial_{r}P|_{r=r_{0}}=-1.79$.}
\end{figure}

Increasing the pressure gradient increases the growth rate for all three classes
of profile, although most dramatically for the `equal drift' case. This
dependence of the growth rate on pressure gradient suggests the $m=2$, $n=1$ DTM
couples to an ideal MHD instability, similar to the interaction between the
$m=1$, $n=1$ kink and tearing modes.\cite{Ara1978a} We verify the presence of
this ideal instability by running a scaling study of growth rate with
resistivity in three sample equilibria. Figure~\ref{fig:ideal_eta}
shows that in the presence of a pressure gradient there is a minimum growth
rate below which $\gamma$ no longer varies with resistivity. This minimum value
increases with increasing pressure gradient, and is not observed in the
force-free equilibrium. While we have not completed the analysis of
this ideal mode, Pritchett et al.\cite
{Pritchett1980}~showed in Cartesian geometry that the DTM tearing layers couple
to a slab-kink mode, the stability of which determines the dependence of the
growth rate on resistivity. We propose that in cylindrical geometry the addition
of a pressure gradient may cause this kink mode to become unstable, thus further
driving the DTM growth. 

We note that Zhao et al.\cite{Zhao2011} found a
similar increase in the DTM growth rate in the presence of a nontrivial pressure
profile. In their simulations, however, the pressure gradient increased the
dependence of the $m=2$, $n=1$ DTM on resistivity~($\gamma \propto \eta^{5/6}$).
The inter-resonance distance of $D=0.26$ examined here is much larger
than than the $D=0.06$ mode considered by Zhao, suggesting that the impact of
the pressure gradient may depend on the spacing between the rational surfaces.

\subsection{Diamagnetic drift effects}
\begin{figure}
	\includegraphics[width=\columnwidth]{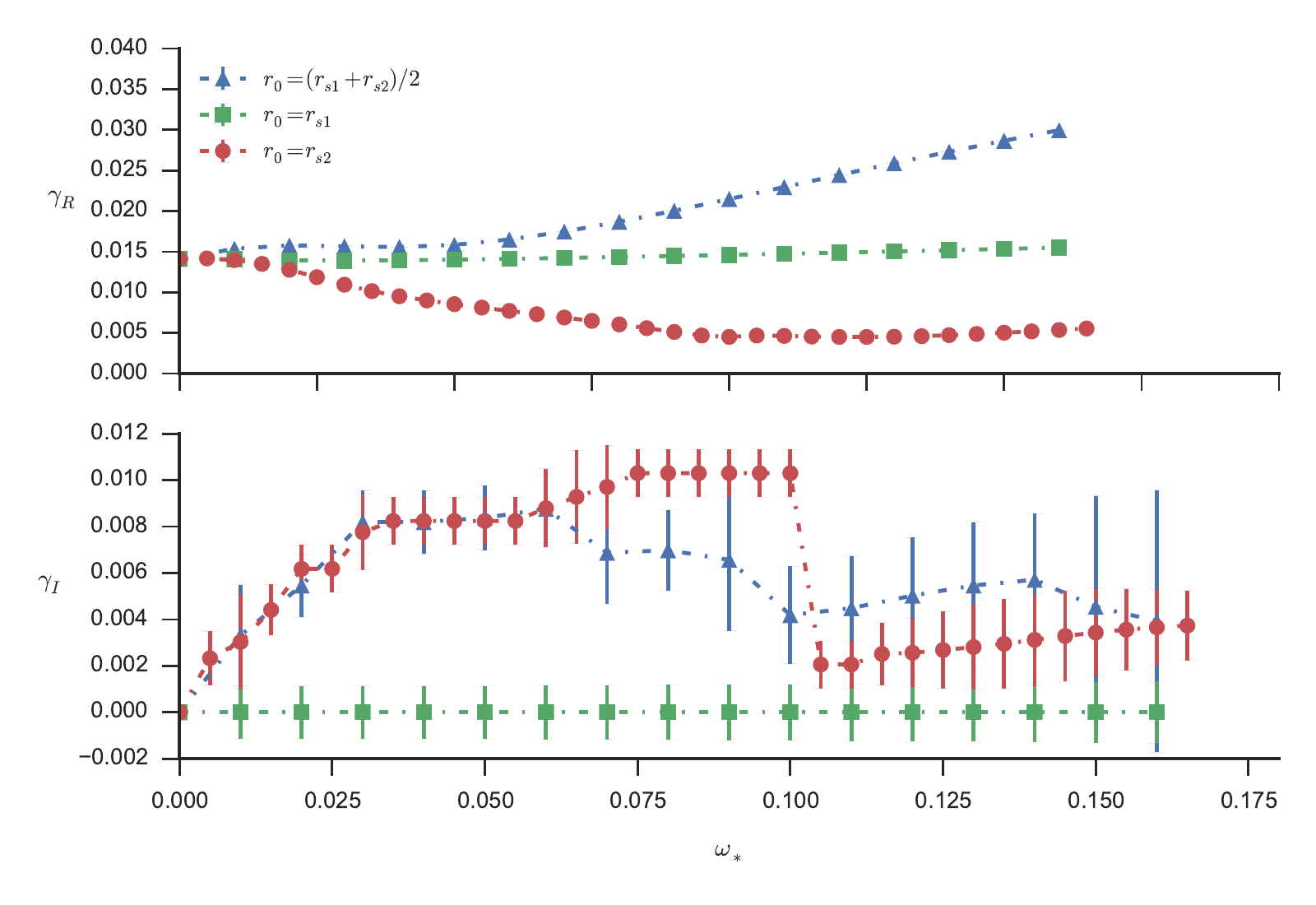}
	\caption{\label{fig:linear_hall}Variation in growth rate $\gamma_{R}$ and
	mode drift frequency $\gamma_{I}$ with
	a diamagnetic drift of $\omega_{*}$ at: both $q=2$ rational surfaces~
	(triangles, equal drift); the $r_{s1}$ surface~(squares, inner drift); the
	$r_{s2}$ surface~(circles, outer drift). Simulations are conducted in Hall MHD,
	$d_ {i}=0.1$, $\rho_{s}=0.014$, with pressure gradient as described in
	Sec.~\ref{sec:equilibrium}}
\end{figure}

Having established via resistive
simulations that the pressure gradient introduces an ideal MHD instability, we
now introduce finite Larmor radius effects and consider how the diamagnetic
drift impacts the cylindrical DTM. We fix the ion inertial length at 
$d_{i}=0.1$ in Eqn.~\ref{eq1:ohmslaw}, 
which results in an ion-sound Larmor radius of 
$\rho_{s} = \sqrt{\beta}d_{i}\approx 0.014$. This large ion
scale is required to provide sufficient scale separation given our use of a
large resistivity to enhance numerical
stability. With $d_{i}$ fixed, increasing the maximum pressure gradient
increases the diamagnetic drift frequency.  In Figure~\ref{fig:linear_hall} we
show the effect of increasing diamagnetic drift at the inner~($r_{0}=r_
{s1}$), outer~($r_{0}=r_{s2}$), or both $q=2$ rational surfaces.

Both the equal drift and inner drift profiles are dominated by the
ideal MHD behavior observed in resistive MHD simulations. For
$\omega_{*}\lesssim0.05$, equal drift at
both $q=2$ rational surfaces counterbalances the enhanced driving energy and
the growth rate remains almost constant with increasing pressure gradient. The
diamagnetic drift is, however, unable overcome the ideal mode at large pressure
gradients and $\gamma_{R}$ again tracks the resistive simulations. We do not
observe any region of constant
growth rate for the inner drift equilibria, and the DTM behavior is dominated by
the ideal MHD driving for all pressure gradients.

\begin{figure}
\centering
\includegraphics[width=\columnwidth]{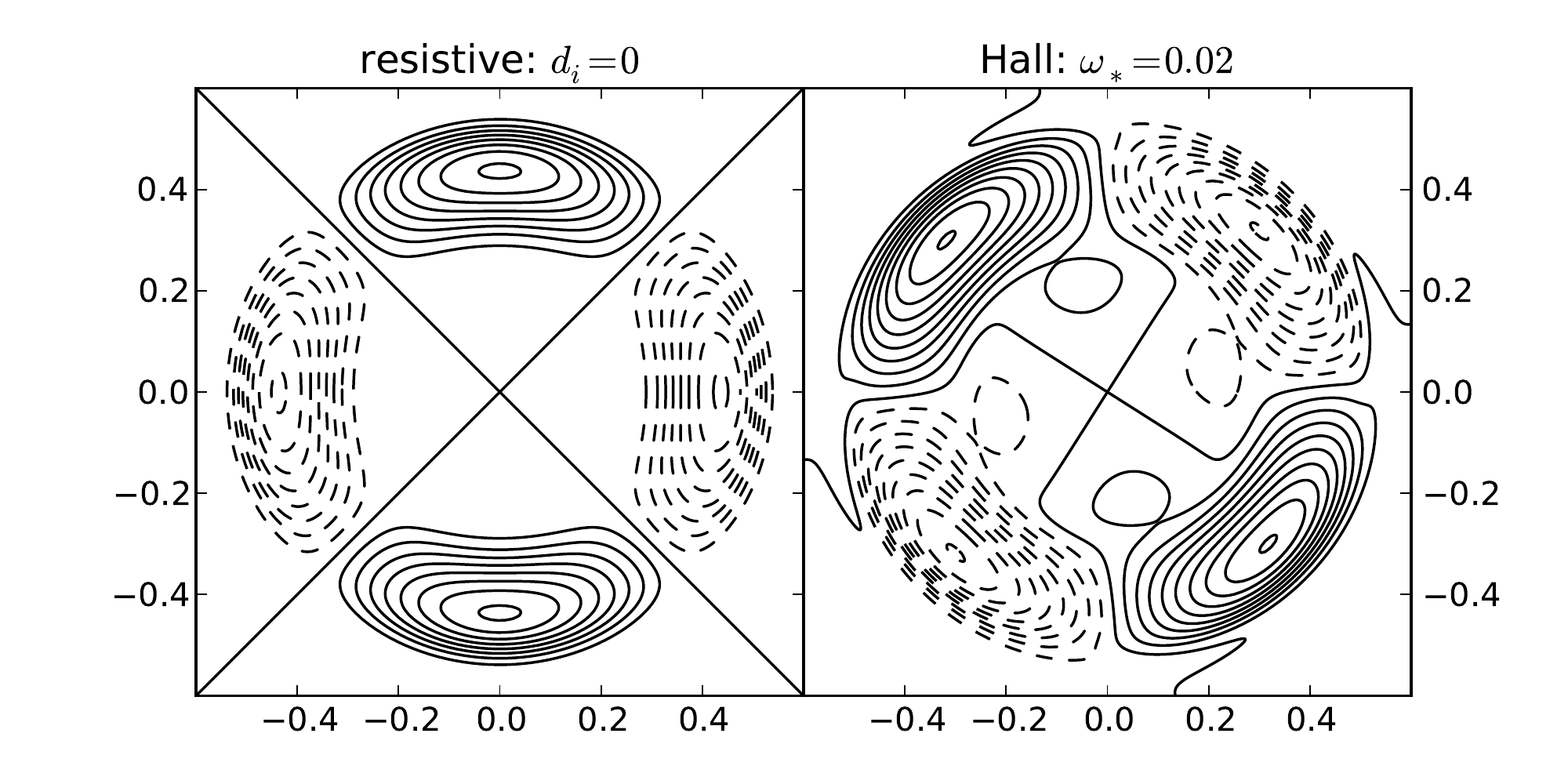}
\caption{\label{fig:linear_sheared_modes}Linear eigenmodes of the helical flux
function $\psi^{*}$ in the presence of pressure gradient centered at $r_{0}=r_
{s2}$ with $N_{b}=0.949$. In restive MHD ($d_{i}=0$, left) there is no
diamagnetic drift; in Hall MHD ($d_{i}=0.1$, right) the outer surface
experiences a drift of $\omega_{*}=0.02$ while the inner surface does not,
resulting in a shearing of the eigenmode.}
\end{figure}

Localized $\omega_{*}$ at the outer resonant surface has a much stronger
stabilizing affect on the DTM. At small drifts ($\omega_{*}\lesssim
0.025$) the growth rate decreases slowly with increasing
$\omega_{*}$. An inflection point is evident in the scaling near $\omega_
{*}=0.025$, after which the growth rate decreases more rapidly and nearly
linearly. The eigenmode at a drift of $\omega_{*}=0.02$ ($r_{0}=r_
{s2}$, $N_{b}=0.949$, $\delta_{N}=0.05$) shows a shearing of the perturbed
helical flux $\psi^{*}$ between the inner and outer $q=2$ surfaces that is
absent in resistive simulations of the same equilibrium~(Figure~\ref
{fig:linear_sheared_modes}). Similar shearing of the eigenmode has been observed
as consequence of equilibrium sheared flow.\cite{Mao2013,Wang2011a}

\begin{figure}
\centering
\includegraphics[width=\columnwidth]{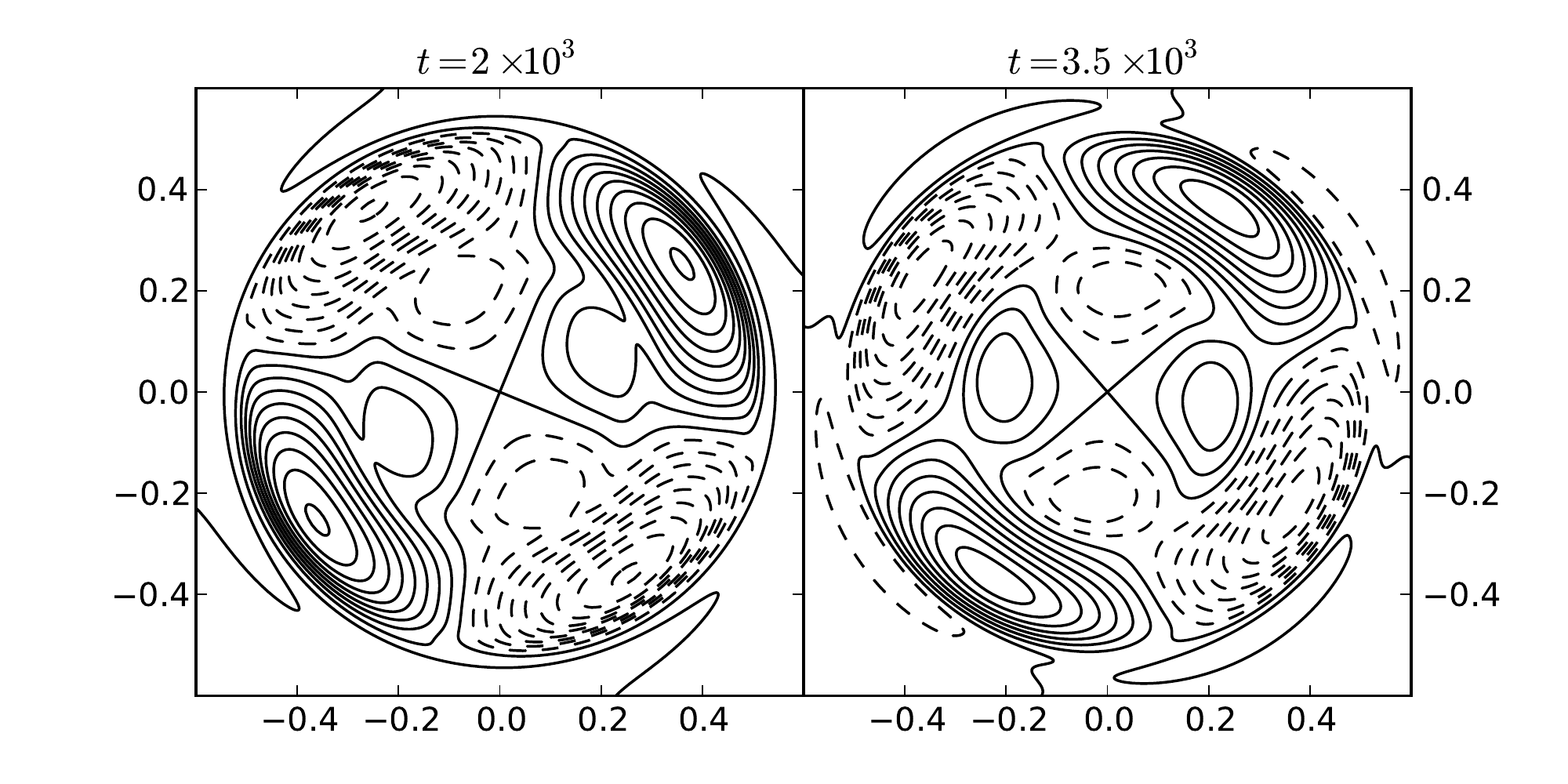}
\caption{\label{fig:linear_decoupled_modes}Two simulation times for a linear
$m=2$, $n=1$ DTM with
pressure profile centered at $r_{0}=r_{s2}$ producing a localized drift of
$\omega_{*}=0.1$. The perturbed helical flux $\psi^{*}$ at the later time
(right) is not a simple rotation of the earlier time (left). Instead the inner
and outer rational surfaces evolve independently, overlapping between the
surfaces. The two tearing layers are decoupled from each
other by the differential diamagnetic drift.}
\end{figure}

Differential diamagnetic drifts above the critical value of $\Delta \omega_
{*}=\omega_{*}(r_{s2})=0.025$ cause decoupling of the reconnecting layers, i.e.
the system acts
predominantly as two independent, drifting, single tearing modes rather than a
single double-tearing mode. Comparing two different times of an $\omega_
{*}=0.1$ outer-drift simulation in Figure~\ref{fig:linear_decoupled_modes}
shows independent movement of the structure around the inner and outer rational
surfaces. This decoupling is, again, similar to that observed in resistive MHD
sheared flow studies.\cite{Mao2013} The continued decrease of the growth rate
above $\omega_{*} \approx 0.25$ is not present in sheared flow equilibria; it is
instead due to stabilizing effects of the diamagnetic drift local to the
singular layer.\cite{Ara1978a,Rogers1995} Thus the outer drift equilibrium
manifests both the decoupling properties of equilibrium sheared flow and the
reconnection inhibiting benefits of the diamagnetic drift.

The scaling behavior in Figure~\ref{fig:linear_hall} shows
significant differences depending on where we apply the diamagnetic drift.
These differences are due to the inherent asymmetry between the two rational
surfaces of the cylindrical DTM. Consider the
growth rate $\gamma_{R}$ and eigenmode drift frequency $\gamma_{I}$ for the
stabilized outer drift~($r_{0}=r_{s2}$) and weakly destabilized inner
drift~($r_{0}=r_{s1}$) profiles. Both feature a strong diamagnetic drift
localized at one of the rational surfaces, producing a differential
$\Delta \omega_ {*}$ that results in decoupling. Increasing $\omega_{*}$
at $r_{0}=r_{s1}$ does not, however, produce any measurable eigenmode drift~
($\gamma_{I}$).
Even in the outer drift ($r_{0}=r_{s2}$) case, where the
perturbation does rotate, only one drifting eigenmode can be found via 
Fourier analysis. This behavior is in contrast to slab-Cartesian sheared flow
studies where two oppositely drifting eigenmodes are observed
post-decoupling.\cite{Mao2013}

The magnetic shear is much greater at the outer rational surface than the inner
(see Fig.~\ref{fig:q_P_ex}), and as a consequence the driving energy local to
the $r_{s2}$ tearing layer is much larger. When the two layers are coupled they
grow as a single mode, but the equilibrium asymmetry causes the
eigenfunction to be biased toward the outer rational surface~(Figure~\ref
{fig:linear_sheared_modes}).
This surface largely dominates the DTM growth. When decoupled
the single layer at $r_{s2}$ is the fastest growing mode; the slower
mode at the inner rational surface cannot easily be detected in our initial
value simulations.

A diamagnetic drift localized near $r_{0}=r_{s1}$ stabilizes and
decouples the weaker inner rational surface. The outer,
dominant surface does not experience any drift, and thus the inner drift
pressure profile does not result in a measurable eigenmode drift~($\gamma_{I}$). 
Nor does the dominant, outer surface encounter
any of the stabilizing effects of the diamagnetic drift. As a consequence,
$\gamma_{R}$ is largely controlled by the destabilization of the ideal MHD mode.

When the drift is instead localized near $r_{0}=r_{s2}$, the outer surface
does rotate and we measure a finite $\gamma_{I}$. The dominant
surface now experiences the stabilizing diamagnetic effects and the growth
rate decreases. For $\omega_{*} > 0.1$ the growth rate again
slowly increases with increasing pressure gradient, and the measured mode drift
frequency $\gamma_{I}$ suddenly drops to a much lower value. Considering
Figure~\ref{fig:linear_decoupled_modes},
the eigenfunction near the inner~($r_{s1}$) singular layer is clearly visible.
The local diamagnetic drift at $r_{s2}$ has thus stabilized
the outer layer sufficiently that it now has a slower growth rate than the
inner, unstabilized tearing mode. Increasing the pressure gradient beyond this
value will have not further decrease the growth rate, as now the ideal MHD
driving of the inner surface has become dominant. 

Our linear simulation results highlight two qualities of this $m=2$, $n=1$
double-tearing mode which limit the stabilizing properties of the diamagnetic
drift. Firstly, the DTM is strongly driven by the interaction between the $q=2$
rational surfaces. Unless a profile provides some differential
rotation effect to decouple the two tearing layers, the diamagnetic drift is not
sufficient to overcome the ideal MHD driving of the increased pressure gradient.
Secondly, the asymmetric magnetic shear inherent to cylindrical geometry causes
one of the $q=2$ rational surfaces to be dominant. For the eigenmode growth rate
to decrease, this fastest growing surface must experience the stabilizing
diamagnetic drift.

\section{Nonlinear Diamagnetic Drifts}
\label{sec:nonlinear}
Based on the linear results of the previous section, we choose the drift of
$\omega_{*}=0.1$ for nonlinear simulations. When localized near the outer
rational surface this diamagnetic drift resulted in the lowest observed growth
rate. To better understand the nonlinear evolution of the DTM we will compare
the $\omega_{*}=0.1$ outer drift profile to both the force-free DTM and equal
and inner drift profiles with the same $\omega_{*}$.

The nonlinear evolution of double-tearing modes is commonly classified by the
growth of the kinetic and perturbed magnetic energies:
\begin{align}
  E_{k} &= \int \frac{1}{2} \rho \mathbf{U} \cdot \mathbf{U} d\mathrm{V} \\
  E_{m} &= -\int \frac{1}{2} \left ( \mathbf{\delta B} \cdot \mathbf{\delta B} +
  2 \mathbf{\delta B} \cdot \mathbf{B}_{0} \right ) d\mathrm{V}
\end{align}
The introduction of diamagnetic drifts in this work will cause the perturbations
to rotate, thus the kinetic energy of all the drifting systems will typically be
larger than the force-free case. As a consequence, the absolute magnitude of
$E_{k}$ is not as good a representation of DTM stability as $E_{m}$. The general
features of the kinetic energy will, however,  provide an indicator of the stage
of DTM evolution.

\begin{figure}
\centering
\includegraphics[width=\columnwidth]{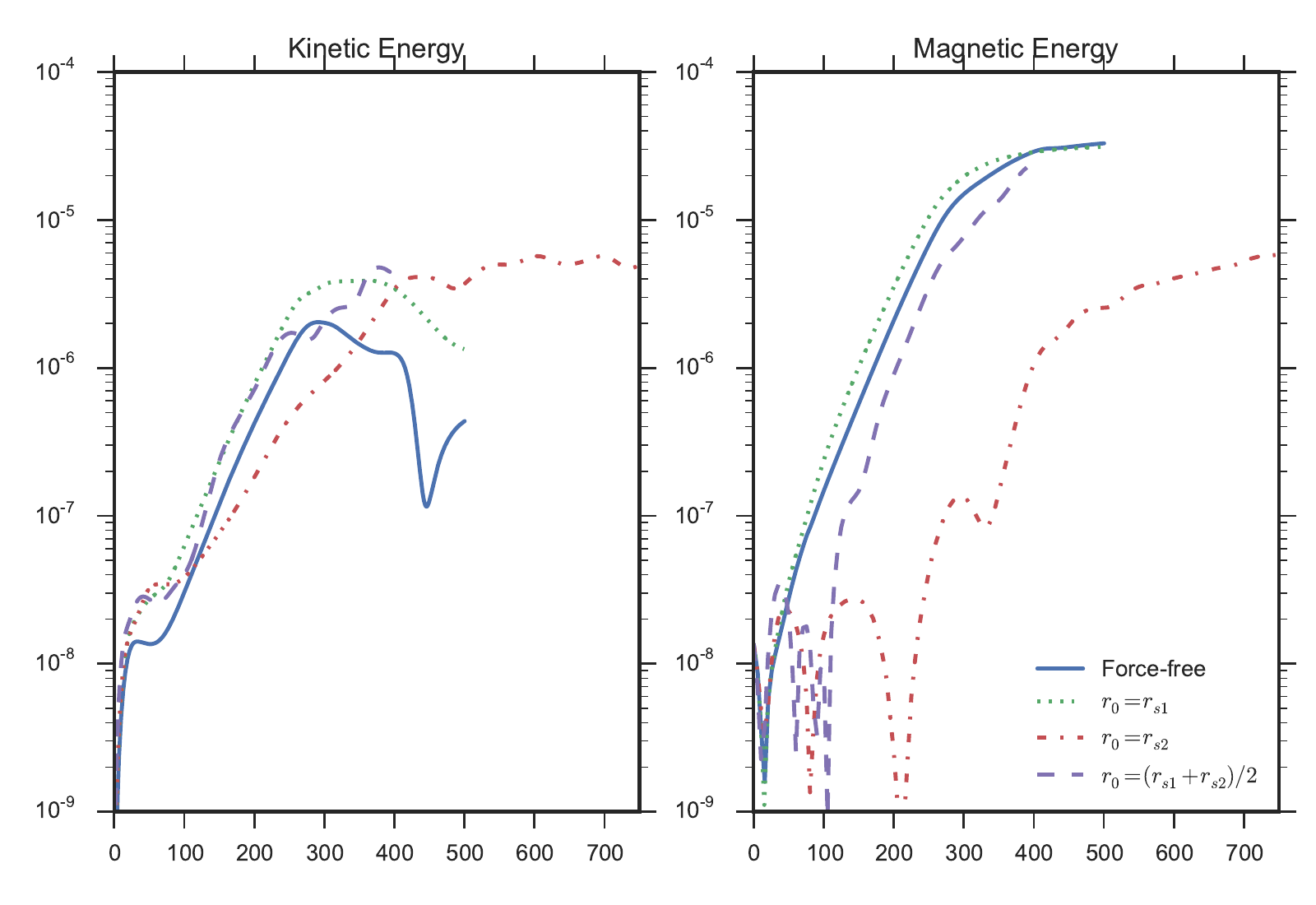}
\caption{\label{fig:ek_em}Nonlinear kinetic and energy growth of the DTM in the
presence of various pressure profiles. The force-free profile has no
equilibrium diamagnetic drift. The remaining profiles have drifts of $\omega_
{*}=0.1$: $r_{0}=(r_{s1}+r_{s2})/2$ - equal drift at both $q=2$ surfaces;
$r_{0}=r_{s1}$ - localized at the inner surface; $r_{0}=r_{s2}$ - localized at
the outer surface.}
\end{figure}

\begin{figure}
\centering
\includegraphics[width=\columnwidth]{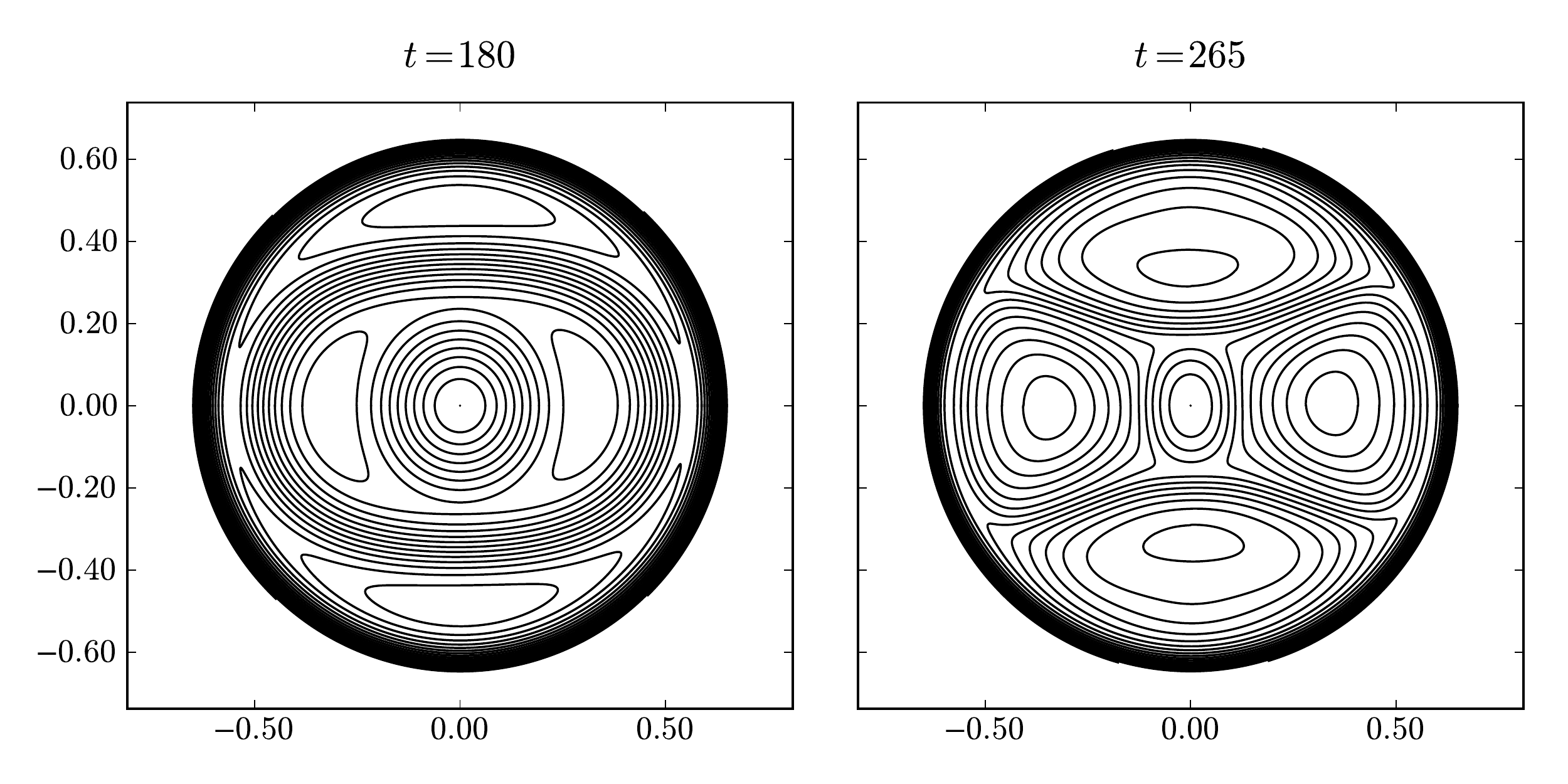}
\caption{\label{fig:force_free_psi} Contours of the helical flux function
$\psi^{*}$ show the island growth regime~(left) and separatrix merging 
event~(right) of the nonlinear, force-free DTM.}
\end{figure}

The kinetic and magnetic energy growth of the force-free, $m=2$, $n=1$ baseline
is shown in Figure~\ref{fig:ek_em}.
The long period of nearly exponential energy growth
represents the development of finite sized magnetic islands such as those 
in Figure~\ref{fig:force_free_psi}. 
In our simulations these magnetic islands do not develop the magnetic structure
necessary for the explosive growth phase observed in higher mode number
DTMs;\cite{Ishii2000,Janvier2011} the kinetic energy of the force-free DTM in
Figure~\ref{fig:ek_em} approaches a maximum value at simulation time
$t\approx265$ smoothly. This maximum
$E_{k}$ corresponds to the separatrix at the inner rational surface merging
with that of the outer, as show in Figure~\ref{fig:force_free_psi}. At this time
the flux between the rational surfaces has been consumed by the reconnecting
layers and the annular current ring is disrupted. Continued
evolution beyond this point results in the magnetic islands reconnecting
completely and relaxation of the system. 

A particular feature of this moderately spaced, low
mode number DTM is that the flux surrounding the magnetic axis is consumed by
reconnection at the $r_{s1}$ rational surface, causing the inner current
sheets to merge across $r=0$. Other simulations of off-axis sawtooth activity in
TFTR\cite{Chang1996} have shown similar behavior, however it is not typically
observed in higher mode number or more closely spaced DTMs.\cite{Ishii2000,Bierwage2005}
In this work we treat the separatrix merging event as a complete loss of system
stability, and thus will not consider whether such highly symmetric behavior
is relevant to realistic devices.

To evaluate the impact of electron diamagnetic drifts nonlinearly we consider
examples from each of the three classes of pressure profile (equal, inner,
and outer drift) that have an equilibrium drift of $\omega_{*}=0.1$ at
both rational surfaces, just the inner, or just the outer. 
To decrease simulation time we seed our 2D helically
symmetric simulations with an $m=2$, $n=1$ perturbation approximately $\num
{1e-4}$ times the equilibrium field, and evolve the system using an
implicit, second order Crank-Nicolson method. For these simulations we use a
grid of 2048 cells in $r$ and 512 in $\theta$, with a nonuniform distribution in
the radial coordinate that increases resolution at and within the $r_{s2}$
surface while decreasing it towards the $r=1$ conducting wall boundary. To aid
numerical stability we choose a resistivity of $\eta=\num{2e-5}$ and set all
other dissipation coefficients to $\num{1e-5}$. We have successfully convergence
tested the following results in both spatial and time resolution, and
also verified that the extra dissipation coefficients do not significantly
impact the mode behavior.

\subsection{Inner drift}
During the linear phase, locating a pressure gradient at the inner rational
surface produced a marginal increase in the growth rate but otherwise did not
significantly impact the mode evolution. Nonlinearly we find that
the dominance of the $r_{s2}$ rational surface persists and a strong
pressure gradient~(and thus drift) at $r_{s1}$ results in slightly faster
growth of the magnetic perturbation when compared to the force-free system~
(Figure~\ref{fig:ek_em}). We find that the island at the
outer rational surface quickly grows large enough to interlock the layer at the
inner surface, which has been previously been observed in nonlinear differential
rotation simulations.\cite{Wang2011} Once recoupled, the separatrix merging event
proceeds with only minor deviation from the force-free system.
Thus locating a drift at the inner, sub-dominant rational surface results in
more system kinetic energy (due to plasma flows near the inner
surface) but is not an effective means of slowing DTM mode growth.

\subsection{Outer rational surface}
Locating a strong diamagnetic drift at the outer resonant surface strongly
stabilizes the nonlinear DTM. The early time growth of $E_{m}$ in Figure~\ref
{fig:ek_em} shows large oscillations, indicating that the two tearing surfaces
are initially decoupled.  At later times these fluctuations continue,
but at a smaller amplitude compared to the total perturbed magnetic
energy.

\begin{figure}
\centering
\includegraphics[width=\columnwidth]{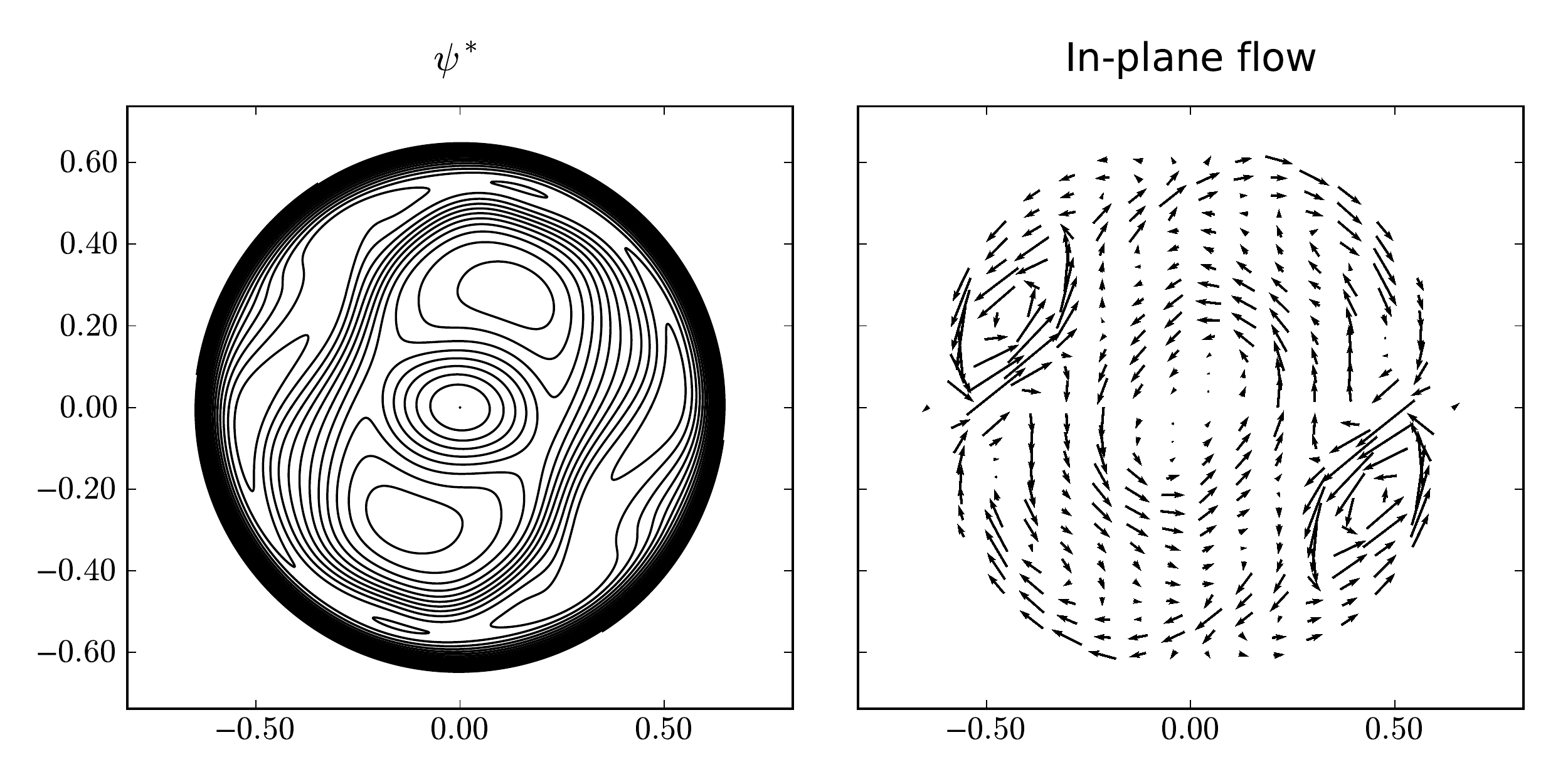}
\caption{\label{fig:outer_final} Applying a diamagnetic drift of $\omega_
{*}=0.1$ at the outer rational surface ($r_{0}=r_{s2}$) saturates the DTM, as
shown by contours of the helical flux $\psi^{*}$~(left) and perpendicular 
flow~(right) at the final simulation time of $t=750$.}
\end{figure}

Considering the state of the system
at the last simulation time ($t=750$, Figure~\ref{fig:outer_final}), 
a significant amount of flux remains between the two $q=2$ surfaces and the
annular current ring is
intact. The plasma flow is largely circulating inside the outer
islands rather than between the two surfaces. The flattening of
the energy growth, together with the relaxed magnetic structure in Figure~\ref
{fig:outer_final}, shows that the fundamental $m=2$, $n=1$ double-tearing mode
is effectively saturated. Intermittent reconnection activity causes late
time fluctuations of $E_{k}$ and $E_{m}$ as the system oscillates, but does
not significantly disturbed the saturated state. At these late times the
source electric field causes slow growth of the perturbed
perturbed magnetic energy, as it drives the system back toward
equilibrium. By comparing to simulations without the source field we
have determined that it has the effect of `pumping' the simulation with energy
rather than allowing it to fully relax.

\subsection{Equal drift}
\begin{figure}
\centering
\includegraphics[width=\columnwidth]{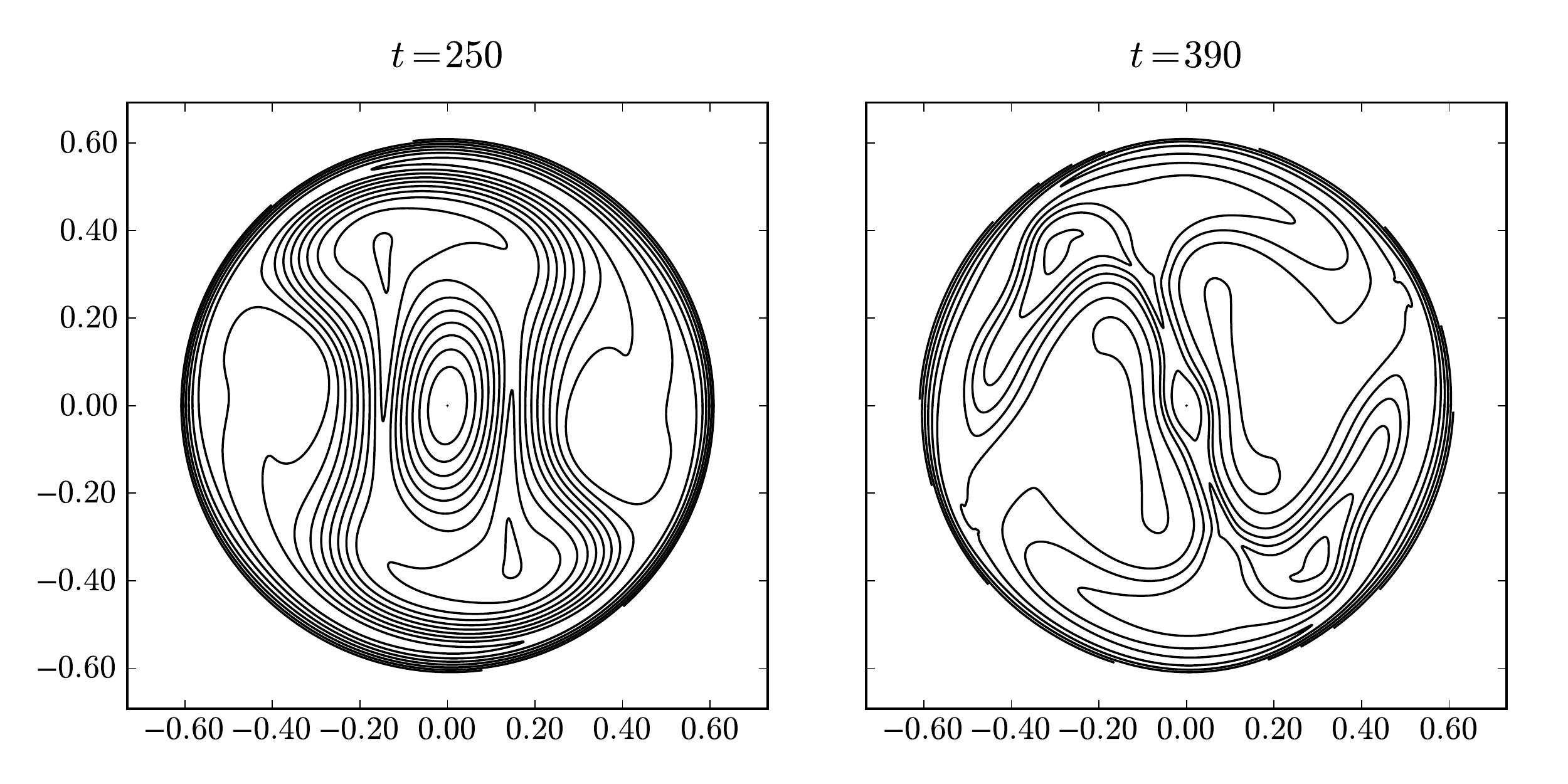}
\caption{\label{fig:equal_cf} Simulation times $t=250$~(kinetic energy
roll-over) and $t=390$~(last simulation time) for an equal drift $\omega_
{*}=0.1$ DTM. Nonlinear enhancement of the pressure gradient severely inhibits
reconnection so that flux remains between the two rational surfaces, but cannot
halt the structural instability.}
\end{figure}

The nonlinear development of the equal drift profile, as viewed through the
energy evolution, is significantly more complicated than the previous case. 
$E_{k}$ growth slows significantly near $t\approx250$ (Figure~\ref {fig:ek_em}),
then rises at a reduced, fluctuating rate. In Figure~\ref{fig:equal_cf} (at
this crest in $E_{k}$) a significant amount of flux remains between the two
tearing surfaces. This roll-over does not, therefore, correspond to a separatrix
merging event. Even at the end of our simulation ($t=390$), when the inner
current sheets approach each other across the magnetic axis, a small amount of
flux remains between the two separatrices. 

\begin{figure}
\centering
\includegraphics[width=\columnwidth]{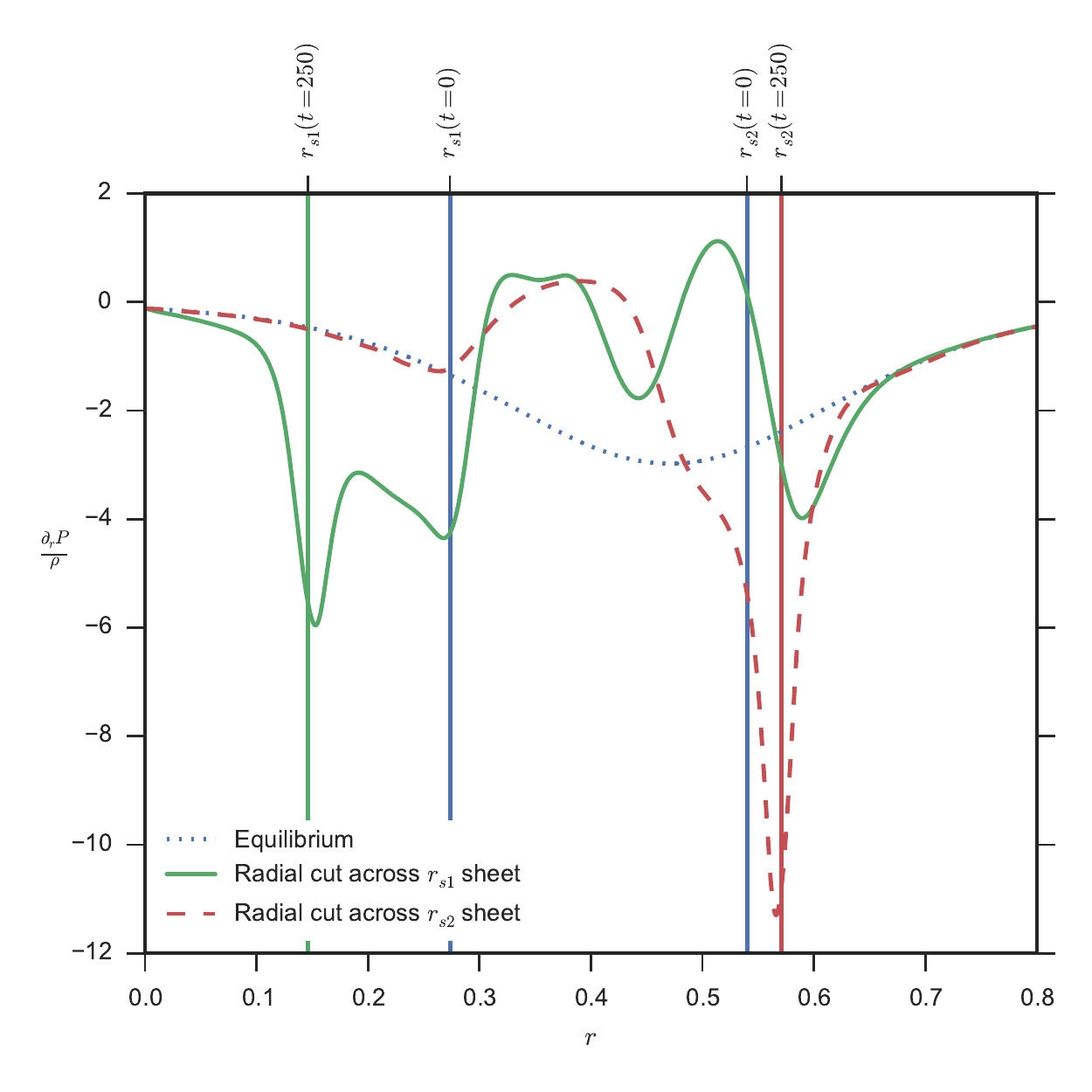}
\caption{\label{fig:equal_gradp} Cuts of $\partial_{r} p/\rho$ across the
inner and outer current sheets at $t=250$ show significant enhancement of
the pressure gradient compared to the equilibrium.}
\end{figure}

The decrease in $E_{k}$ and $E_{m}$ growth is a result of
nonlinear evolution of the pressure profile. Figure~\ref{fig:equal_gradp} shows
cuts of $\partial_{r}p/\rho$ (the pressure and density
contribution to Eqn.~\ref{eq:drift}) across the inner and outer current sheets
at time $t=250$ (the roll-over point) compared to the equilibrium profile. The
nonlinear growth of the magnetic islands distorts the equilibrium pressure
gradient and results in a significant enhancement to the diamagnetic drift
within the tearing layers. As a consequence, reconnection is highly suppressed. 

Similar nonlinear enhancement of the pressure gradient has been proposed as a
mechanism for saturation $m=1$ kink-tearing mode.\cite{Rogers1995,Beidler2011} In
our DTM simulation, however, this nonlinear enhancement of $\omega_{*}$ does
not lead to complete stabilization. The magnetic islands continue to evolve, and
intermittent reconnection occurs. This continued growth of the kinetic and
magnetic energies~(Fig.~\ref{fig:ek_em}), and deformation of the 
separatrices~(Fig.~\ref{fig:equal_cf}) is a consequence of the large
islands sizes required to enhance the pressure gradient sufficiently to
cut off reconnection. The magnetic structure has
become unstable, similar to previous simulations of explosive-type
double-tearing modes,\cite{Ishii2002,Janvier2011} and thus the instability
continues to develop instead of saturating. In this respect the
equal drift profile is less desirable than the outer drift, which stabilized the
mode before significant deformation.

\section{Conclusion}
\label{sec:conclusion}
In this work we have shown that electron diamagnetic drifts can have a
stabilizing effect on the $m=2$, $n=1$ double-tearing mode. Their efficacy
depends, however, on where they are located. Linearly we were only able to
achieve a substantial decrease in the DTM growth rate by localizing a strong
pressure gradient at the outer rational surface. This class of profile combined
the decoupling properties of differential plasma rotation with the local
stabilizing
benefits of the diamagnetic drift on the dominant tearing layer. The equal and
inner drift profiles contained only one of these stabilizing effects, and thus
could not overcome the increased ideal MHD driving caused by the pressure
gradient.

Nonlinearly we found the decoupling and local stabilization effects of the outer
drift profile resulted in saturation of the DTM at finite amplitude. The
preservation of the annular current ring in this simulation indicates that
diamagnetic drifts may act as a mechanism for stabilizing off-axis sawtooth
crashes. We also found that nonlinear enhancement of the pressure gradient
in an equal drift profile was able to significantly slow the growth of the
instability. The large islands necessary to halt reconnection, however, resulted
in unstable magnetic structure. 

Our work shows that when finite Larmor radius effects are included
the presence of DTM activity depends strongly on the details of the plasma
pressure. The equilibria considered in this work are highly constrained, and
more realistic profiles will likely include some intermediate form of our
results. Whether such profiles are a viable mechanism for avoiding DTM
driven off-axis sawtooth behavior in RMS devices is currently unclear.
Extending these results to high mode number DTMs and toroidal geometry would
allow better comparison to experimental data, as would consideration of
non-constant temperatures and hot ions. We have provided strong evidence, 
however, for the dependence of DTM activity on the location of steep
pressure profiles in the plasma.

\begin{acknowledgments}
This work was supported by the U.S.~Department of Energy, Office of Science, 
Office of Fusion Energy Sciences under Award Number DESC0006670. It contains
research from a dissertation submitted to the Graduate School at 
the University of New Hampshire as part of the requirements for completion of
Stephen Abbott's doctoral degree in Physics.

Calculations were performed using: Trillian, a Cray
XE6m-200 supercomputer at UNH supported by the NSF MRI program under grant
PHY-1229408; and Fishercat, an IBM Blade Center H supported by the NSF CRI
program under grant CNS-0855145.

\end{acknowledgments}
\bibliography{cylindricaldrift}
\end{document}